\documentclass[dvips]{article}

\usepackage{icrc2011}
\newcommand{\lsi}{LS~I~+61~303}

\def\apj{ApJ}%
\def\apjl{ApJL}%
\def\aap{A \& A}%
\def\mnras{MNRAS}%
\title{Detection of \lsi\ in a low VHE $\gamma$-ray emission state with the MAGIC telescopes}

\shorttitle{T. Jogler et~al. \lsi\ in a low VHE $\gamma$-ray
emission}

\authors{T. Jogler$^{1}$, O. Blanch$^{2}$ for the MAGIC collaboration}
\afiliations{$^1$Max-Planck-Institut f\"ur Physik, D-80805
M\"unchen, Germany\\ $^2$IFAE, Edifici Cn., Campus UAB, E-08193
Bellaterra, Spain}
\email{jogler@mppmu.mpg.de}

\abstract{The gamma-ray binary system \lsi\ was studied in great
detail in VHE gamma-rays in the last years by the MAGIC telescope.
The VHE emission of the system exhibited a prominent periodic
outburst in the orbital phases 0.6-0.7 between September 2005 to
January 2008. In Fall 2008 the Fermi collaboration reported as
well periodic emission in the MeV to GeV energy range, but with a
shifted outburst in the phases 0.35-0.45. MAGIC observed again LS
I+61 303 in 2009 with the twice more sensitive stereo mode to
allow for detailed correlation studies between the VHE gamma-ray
and Fermi/LAT energy band. Here we present our new results, which
show a significant reduction in the VHE gamma-ray flux in the
phase of the periodic outburst by almost one order of magnitude
compared to our previous measurements. Furthermore, the 0.1-phase
averaged light curve shows no significant outburst, but a rather
constant flux. Here we will discuss the implications of our
results for future gamma-ray studies of \lsi. }
\keywords{ gamma-rays, binaries, observation, \lsi\, MAGIC}

\begin{document}
\maketitle


\section{Introduction}

The \lsi\ system consists of a Be star and a compact object of
still uncertain nature, either a neutron star or a black hole. Its
orbital period, which is most precise measured in radio, is
$26.4960\pm0.0028$ days~\cite{Gregory:2002}. Soft X-ray outbursts
modulated with the same period as in the radio waveband were
reported by~\cite{Paredes_1997A&A...320L..25P} but the peak of the
emission changes its phase over several
years~\cite{Torres_2010ApJ...719L.104T}. Many orbital parameters
of the system remain controversial (see
\cite{Casares:2005wn,Grundstrom:2006,Argona:2009}) but
observations indicate a highly eccentric orbit ($e=0.55\pm0.05$)
with the periastron passage at orbital phase
$\phi_{\mathrm{per}}=0.275$~\cite{Argona:2009}. These orbital
parameters are important for modelling the VHE emission of the
system as shown in,e.g.,
\cite{Sierpowska-Bartosik:2009ApJ...693.1462S} or
\cite{Dubus_rel_boosting_2010A&A...516A..18D}.

In 2006 the MAGIC collaboration discovered variable VHE
$\gamma$-ray emission from \lsi~\cite{MAGIC_lsi_science:2006vk}. A
following  extensive observational campaign in Fall 2006 found a
period for the VHE emission of $26.6\pm0.2$
days~\cite{MAGIC_lsi_periodic:2009ApJ...693..303A}. The VHE
$\gamma$-ray emission showed an outburst in the orbital phase
interval 0.6--0.7 with no significant $\gamma$-ray emission
detected during the rest of the orbit. In particular, no VHE
$\gamma$-ray signal was detected by MAGIC around the periastron
passage of the system. The data from Fall 2006 also suggested a
correlation between the X-ray and VHE $\gamma$-ray flux. An
extensive multi-wavelength campaign conducted in 2007, including
MAGIC, XMM-Newton and Swift, provided strong evidence for the
X-ray/VHE $\gamma$-ray flux correlation in the strictly
simultaneous taken
data~\cite{MAGIC_lsi_xrayvhe:2009ApJ...706L..27A}. In contrast, no
correlation was found between the radio wavelength flux and the
VHE $\gamma$-ray flux from the Fall 2006
campaign~\cite{MAGIC_2008ApJ...684.1351A}.

The VHE emission of \lsi\ was confirmed by
VERITAS~\cite{Veritas_lsi_discovery:2008ApJ...679.1427A}. However,
in observations conducted by the VERITAS collaboration in Fall
2008 and early 2009, no VHE signal was detected. More recent
VERITAS observations in Fall 2009 (the same time period considered
in the present paper) also yielded only upper limits for the VHE
emission from \lsi~\cite{VERITAS_lsi_2011arXiv1105.0449A}. Very
recently the VERITAS collaboration reported a detection of the
system with a significance of more than
$5\sigma$~(\cite{2010ATel_Veritas,VERITAS_lsi_2011arXiv1105.0449A})
between orbital phases 0.05 and 0.23. This places the detection at
superior conjunction and 5.8 to 1.3 days before the periastron
passage a phase range where no VHE $\gamma$-ray emission was
previously detected.

The binary system was observed in high energy (HE, $0.1-
100\mathrm{ GeV}$) $\gamma$-rays by
EGRET~\cite{EGRET_disc_1997ApJ...486..126K}, but the large
position uncertainty of the source and inconclusive variability
studies of the emission\cite{Tavani}, prevented its unambiguous
identification. The spacial association with \lsi\ was only
achieved following HE $\gamma$-ray observations by
AGILE~\cite{AGILE_discovery_2009A&A...506.1563P}. Most recently,
Fermi/LAT found that the HE $\gamma$-rays are periodically
modulated in very good agreement with the (radio) orbital
period~\cite{FERMI_disvoery_2009ApJ...701L.123A} and firmly
established \lsi\ as the counterpart of the HE emission. The HE
outburst was not, however, observed at the same phases as the VHE
outburst but occurred just after the periastron passage between
phases 0.3--0.45. This difference in phase may indicate that
different processes are responsible for the HE and VHE
$\gamma$-ray emission. On the other hand, the same process might
produce both emissions if the GeV $\gamma$-rays are produced by
inverse Compton (IC) pairs cascading developing in the radiation
field of the star. For more details on such a scenario
see~\cite{2006MNRAS.368..579B}. Another possibility is that the
shift in the peak emission could be caused by a different location
of the $\gamma$-ray production site in the
system~\cite{2011A&A...527A...9Z}. We note that no simultaneous
VHE observations are available for the same time (Aug 2009 to Jan
2009) of the first reported Fermi observations
\cite{FERMI_disvoery_2009ApJ...701L.123A}. An unambiguous
interpretation of the non simultaneous SED from MeV to TeV
energies of \lsi\ is not possible because the system might have
changed its VHE emission in the meantime.

Two principal scenarios have been proposed to explain the
nonthermal emission from \lsi: an accretion powered microquasar
(e.g.
\cite{Romero_2005ApJ...632.1093R,Bednarek:2006,Gupta_2006ApJ...650L.123G,Bosch-Ramon_2006A&A...459L..25B})
and a rotational powered compact pulsar wind (e.g.
\cite{Dubus:2006,Sierpowska-Bartosik:2009ApJ...693.1462S,Zdziarski_2010MNRAS.403.1873Z}).
An alternative model assumes that the compact object is an
accreting magnetar and that the  $\gamma$-rays are produced along
the accretion flow onto the magnetar~\cite{2009MNRAS.397.1420B}.
Although high resolution radio measurements
\cite{Dhawan_2006smqw.confE..52D}
 show an extended and variable structure emerging from the
system, it is not clear if they are produced by collimated plasma
outflows (jets) or by a pulsar wind interacting with that of the
Be star. Neither of the two proposed scenarios could be validated
by accretion disk features, e.g. a thermal component in the X-ray
spectrum, or the presence of pulsed emission at any wavelength.
Thus the engine behind the VHE emission remains an open question.

Here we present new observations of \lsi\ conducted with the MAGIC
stereo system. This has twice the sensitivity of the previous
MAGIC campaigns, and results in a significant detection of the
binary system during a newly identified low emission state.

\section{Observations}

The observations were performed between 2009 Oct 15 and 2010 Jan
22 using the MAGIC telescopes on the Canary island of La Palma
($28.75^\circ$N, $17.86^\circ$W, 2225~m a.s.l.), from where \lsi\
is observable at zenith distances above 32$^{\circ}$. The MAGIC
stereo system consists of two imaging air Cherenkov telescopes,
each with a 17~m diameter mirror. The observations were carried
out in stereo mode, meaning that only shower images which trigger
simultaneously both telescopes are recorded. The stereoscopic
observations provides a $5\sigma$ signal above 300 GeV from a
source which exhibits 0.8\% of the Crab Nebula flux in 50 hours
observation time, a factor of two more sensitive than our single
telescope campaign on \lsi\ in 2007. Further details on the design
and performance of the MAGIC stereo system can be found
in~\cite{MAGIC_stereo_performance}.

The \lsi\ data set spans four orbits of the system, with two of
these orbits observed for only one and three nights, respectively.
The data taken in 2009 Oct and 2009 Nov were restricted to
moonless nights. The data sample the orbital phases 0.55 to 0.975
for 2009 Oct , and 0.575 to 1.025 for 2009 Nov , the last night of
which is in the next orbital cycle. The data recorded in 2010 Jan
cover the phases 0.22 to 0.32 and were obtained during moonlight
conditions. All data were taken at zenith angles between
32$^{\circ}$ and 48$^{\circ}$. After pre-selection of good quality
data a total of 48.4~hours of data remained for the analysis. The
observation strategy aimed to cover consecutive nights with at
least three hours of observation in each individual night. Due to
adverse observation conditions such as bad weather or too high
zenith angle, the data set does not have uniform coverage during
the orbital phases and some nights have shorter observation times
than the planned three hours.

\section{Data Analysis}

The data analysis was performed with the standard MAGIC
reconstruction software. The recorded shower images were
calibrated, cleaned and used to calculate image parameters
individually for each telescope. The energy of each event was then
estimated using look up tables generated by Monte Carlo (MC)
simulated $\gamma$-ray events. Events that triggered both
telescopes simultaneously were then selected (so called stereo
events)\footnote{This step is only needed for the 2009 October
data where no hardware stereo trigger was yet available.} and
further parameters, e.g. the height of the shower maximum and the
impact parameter from each telescope, were calculated. The gamma
hadron classifications and reconstructions of the incoming
direction of the primary shower particles were then  performed
using the Random Forest (RF) method~\cite{magic:RF}. Finally, the
signal selection used cuts in the hadronness (calculated by the
RF) and the squared angular distance between the shower pointing
direction and the source position  ($\theta^2$). The energy
dependent cut values were determined by optimizing them on a
sample of events recorded from the Crab Nebula under the same
zenith angle range and similar epochs than the \lsi\ data. For the
energy spectrum and flux, the effective detector area was
estimated by applying the same cuts used on the data sample to a
sample of MC simulated $\gamma$-rays. Finally, the energy spectrum
was unfolded, accounting for the energy resolution and possible
energy reconstruction bias~\cite{magic:unfolding}.

\section{Results}

\begin{figure}[tbp]
  \centering
  \includegraphics[width=\linewidth]{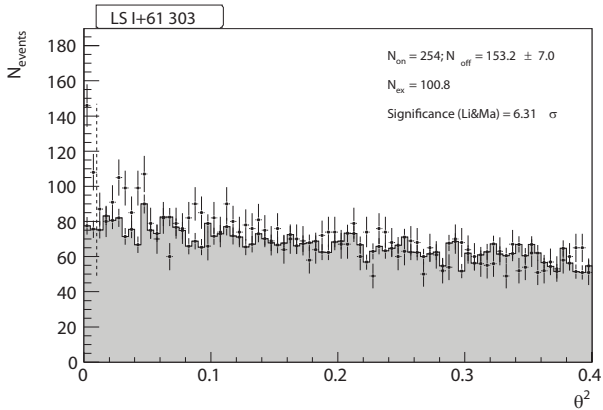}
  \caption{The squared angular distance between the pointing
direction of the shower and the source position  ($\theta^2$-plot)
for the position of \lsi\ (points) and the simultaneous determined
background regions (grey shaded histogram) for the total MAGIC
data set.
  }
  \label{fig:detection}
\end{figure}

The integral data set of 48.8 hours shows a $6.3 \sigma$ detection
of VHE $\gamma$-ray emission above 400 GeV from \lsi\ (see
Fig.~\ref{fig:detection}). The integrated flux above 300 GeV is
\begin{equation}
F(E>300 \mathrm{GeV}) =  (1.4 \pm 0.3_\mathrm{st} \pm
0.4_\mathrm{sy}) \times 10^{-12} \mathrm{cm}^{-2} \mathrm{s}^{-1}.
\end{equation}
corresponding to about 1.25\% of the Crab Nebula flux in the same
energy range.

\subsection{Light curve}

\begin{figure}[tb!]
  \centering
  \includegraphics[width=\linewidth]{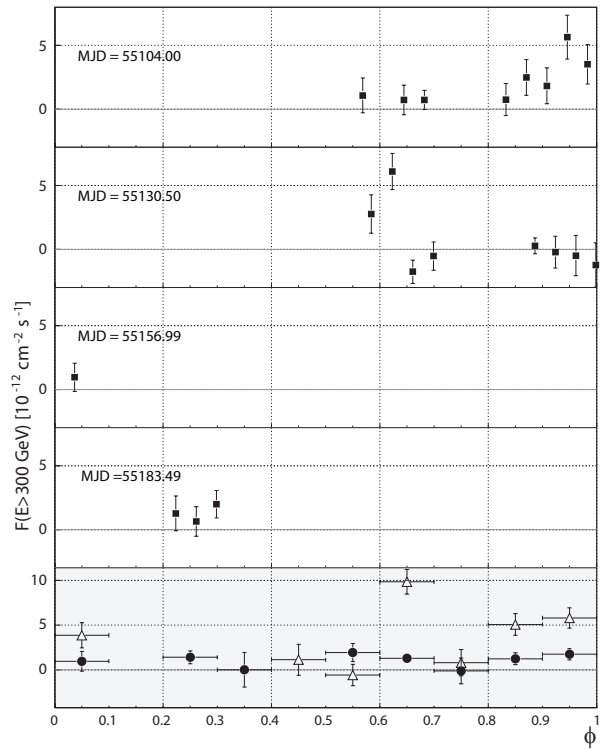}
  \caption{VHE ($E>300\mathrm{GeV}$) $\gamma$-ray flux of \lsi\ as a
    function of the orbital phase for the four observed orbital cycles
    (four upper panels) and averaged for the entire observation time
    (lowermost panel, black points). The starting MjD for each orbital cycle is given in
    each panel. In the lower most panel the previous published
     \cite{MAGIC_lsi_xrayvhe:2009ApJ...706L..27A} averaged fluxes per phasebin
     are shown as hollow triangles too. Vertical error bars show $1\sigma$ statistical
    errors.}
  \label{fig:lc}
\end{figure}

We derived a nightly light curve above an energy of 300~GeV that
is shown in Fig.~\ref{fig:lc}. A constant flux fit to the light
curve yields a $\chi^2 / \mathrm{dof}=42.15 / 19 $ ($p=1.5\times
10^{-3}$) and hence is unlikely. Thus, as in previous
observations, the emission is variable and reaches a maximum flux
around orbital phase 0.62 of $F(E>300 \mathrm{GeV}) = (6.1
\pm1.4_\mathrm{st} \pm 1.8_\mathrm{sy}) \times 10^{-12}
\mathrm{cm}^{-2} \mathrm{s}^{-1}$ corresponding to 5.4\% of the
Crab Nebula flux. This is a much lower peak emission than detected
in our previous campaigns at the same orbital phases and sampled
with very similar cadence. For a more quantitative comparison of
the 2009 emission level with the previous MAGIC observations, we
included the phase averaged 2007 light curve in Fig.~\ref{fig:lc}.

It is clear that the averaged emission level is lower than the
data from 2005 to 2007. Not only had the flux changed but a VHE
$\gamma$-ray excess was also observed at phases other than those
of the periodic outburst between 0.6--0.7. The highest flux is,
however, again detected in an outburst during the interval
0.6--0.7, and the measurements in the orbital cycle of 2009 Nov
show the same burst profile as in previous observations but with a
reduced flux level. Whether this outburst is still a periodic
property of the VHE emission for \lsi\ cannot be determined with
the small sample of orbital cycles observed in this campaign; it
is noteworthy that the outburst was not detected during the orbit
observed in 2009 Oct.

The rather low mean values, even in the phase bin 0.6--0.7, of the
individual night peak emission indicates that most of the emission
of the system is contributed by only few nights. This is the case
at least for the phase bins 0.6--0.7 and 0.9--1.0, where the mean
flux shows the highest significance. On the other hand, there are
similar fluxes, measured in phase intervals 0.2--0.3 and 0.5--0.6,
although with slightly lower significances, without any obvious
individual high flux nights.

\subsection{Spectrum}

The emission level of \lsi\ was too low during most phases to
obtain statistical significant phase dependent spectra. The total
signal, however, was sufficient to form a phase integrated
spectrum for the source. The spectrum is well described by a
simple power law
\begin{equation}
\frac{\mathrm{d}F}{\mathrm{d}E} =
\frac{(2.3\pm0.6_\mathrm{st}\pm0.7_\mathrm{sy}) \cdot 10^{-13}}
{\mathrm{TeV}\,\mathrm{cm}^2\,\mathrm{s}} \frac{E}{1\,
\mathrm{TeV}}^{-2.5\pm0.5_\mathrm{st}\pm0.2_\mathrm{sy}} \, ,
\end{equation}
with a reduced $\chi^2 = 0.24/2$. The spectral fit parameters
match those previously reported by
MAGIC~\cite{MAGIC_lsi_science:2006vk,MAGIC_lsi_periodic:2009ApJ...693..303A,MAGIC_lsi_xrayvhe:2009ApJ...706L..27A}
showing no evidence for long term spectral variability despite
very different fluxes during these different campaigns.

\section{Conclusion}

The binary system \lsi\ was detected emitting VHE $\gamma$-rays in
2009 at a level a factor 10 lower than previously observed in the
phase interval 0.6--0.7. The previously observed orbital modulated
outburst in this interval was not detected during the first
observed orbital cycle, whereas it was observed in the second.
From this data set alone it was not possible to determine whether
the outburst is still a truly periodic feature of the light curve.
Although we could not significantly detect emission at individual
orbital phases in the here presented observations because of the
weakness of the source, it appears that several phase intervals
dominate in the integral signal. This is strong evidence for a new
behavior in the VHE $\gamma$-ray emission of \lsi. In previous
observations conducted with MAGIC, the 0.6--0.7 interval dominated
the total flux.

Furthermore it is evident that the flux during the phase interval
0.6--0.7 is considerable reduced compared to the previous
campaigns, and on a similar level as in other phase intervals
(e.g. 0.9--1.0). This suggests that a change in the VHE
$\gamma$-ray emission of \lsi\ has occurred. On the other hand,
there was no statistically significant  change in the spectrum of
the orbit-integrated flux in 2009 compared to the earlier results,
suggesting that  the same processes continue to produce VHE
gamma-rays, but that either fewer are produced or they are more
absorbed.

If enhanced absorption decreases the observed VHE flux, the photon
energy will be redistributed to lower energies and thus might be
visible as a flux increase at lower energies. The produced VHE
$\gamma$-rays would thus need to propagate through a circum-source
environment denser in photons and we would expect to detect a
cut-off or an absorption feature in our spectrum, neither of which
is found. Despite the richness of observations of the system, its
VHE emission is still not understood.

The VERITAS observations in the same period as the one considered
here did not detect VHE $\gamma$-ray emission from the
system~\cite{VERITAS_lsi_2011arXiv1105.0449A}. Our measurements
are not, however, in contradiction to those of VERITAS. Our longer
integration combined with a denser sampling of two orbital cycle
yielded a fainter detection threshold than the expected VHE
$\gamma$-ray signal from \lsi\ from previous campaigns expected.
Thus it is evident that a frequent sampling with long individual
integrations is required not to miss weak emission from binary
systems.

This is the first VHE $\gamma$-ray detection of \lsi\ in the era
of the Fermi satellite. The faint emission at VHE $\gamma$-rays
does not yet permit night by night correlation studies but do show
that the emission in \lsi\ has changed on longer timescale since
2007. More sensitive and even deeper VHE $\gamma$-ray observations
should yield shorter timescale correlation studies.

\section*{Acknowledgments}

We would like to thank the Instituto de Astrof\'{\i}sica de
Canarias for the excellent working conditions at the Observatorio
del Roque de los Muchachos in La Palma. The support of the German
BMBF and MPG, the Italian INFN, the Swiss National Fund SNF, and
the Spanish MICINN is gratefully acknowledged. This work was also
supported by the Marie Curie program, by the CPAN CSD2007-00042
and MultiDark CSD2009-00064 projects of the Spanish
Consolider-Ingenio 2010 programme, by grant DO02-353 of the
Bulgarian NSF, by grant 127740 of the Academy of Finland, by the
YIP of the Helmholtz Gemeinschaft, by the DFG Cluster of
Excellence ``Origin and Structure of the Universe'', and by the
Polish MNiSzW Grant N N203 390834.



\begin{thebibliography}{}
\bibitem{Gregory:2002}Gregory, P.~C.,\apj, 2002 {\bf 575}, 427

\bibitem{Paredes_1997A&A...320L..25P}
{Paredes}, J.~M., {Marti}, J., {Peracaula}, M., and {Ribo}, M.,
\aap, 1997, {\bf 320}, L25

\bibitem{Torres_2010ApJ...719L.104T}
{Torres}, D.~F. et~al.,\apjl, 2010, {\bf 719}, L104

\bibitem{Argona:2009}{Aragona}, C. et~al., \apj , 2009,{\bf 698}, 514

\bibitem{Grundstrom:2006}Grundstrom, E.~D. et~al.,Astrophys. J. , 2007,{\bf 656}, 437

\bibitem{Casares:2005wn}
Casares, J., Ribas, I., Paredes, J.~M., Marti, J., and
Allende~Prieto, C.,\mnras, 2005, {\bf 360}, 1091

\bibitem{Sierpowska-Bartosik:2009ApJ...693.1462S}
{Sierpowska-Bartosik}, A. and {Torres}, D.~F.,\apj, 2009, {\bf
693}, 1462

\bibitem{Dubus_rel_boosting_2010A&A...516A..18D}
{Dubus}, G., {Cerutti}, B., and {Henri}, G.,\aap, 2010 {\bf 516},
A18+

\bibitem{FERMI_disvoery_2009ApJ...701L.123A} {Abdo}, A.~A.
et~al., \apjl, 2009 {\bf 701}, L123

\bibitem{Veritas_lsi_discovery:2008ApJ...679.1427A}
{Acciari}, V.~A. et~al., \apj, 2008 {\bf 679}, 1427

\bibitem{VERITAS_lsi_2011arXiv1105.0449A}
{Acciari}, V.~A. et~al., ArXiv e-prints, 2011

\bibitem{MAGIC_lsi_science:2006vk}
Albert, J. et~al., Science, 2006, {\bf 312}, 1771

\bibitem{magic:unfolding}
Albert, J. et~al., NIM., 2007,{\bf A583}, 494

\bibitem{magic:RF}
Albert, J. et~al., NIM., 2008, {\bf A588}, 424

\bibitem{MAGIC_2008ApJ...684.1351A}
{Albert}, J. et~al., \apj , 2008, {\bf 684}, 1351

\bibitem{MAGIC_lsi_periodic:2009ApJ...693..303A}
{Albert}, J. et~al., \apj, 2009, {\bf 693}, 303

\bibitem{MAGIC_stereo_performance}{Aleksi\'{c}}, J. et~al.,
2011, ArXiv e-prints, 2011

\bibitem{MAGIC_lsi_xrayvhe:2009ApJ...706L..27A}
{Anderhub}, H. et~al., \apjl, 2009,{\bf 706}, L27

\bibitem{Bednarek:2006}{Bednarek}, W.,
\mnras, 2006a, {\bf 371}, 1737

\bibitem{2006MNRAS.368..579B}{Bednarek}, W.,
\mnras, 2006b, {\bf 368}, 579

\bibitem{2009MNRAS.397.1420B}
{Bednarek}, W., \mnras , 2009, {\bf 397}, 1420

\bibitem{Bosch-Ramon_2006A&A...459L..25B}
{Bosch-Ramon}, V., {Paredes}, J.~M., {Romero}, G.~E., and
{Rib{\'o}}, M., \aap , 2006, {\bf 459}, L25


\bibitem{Dhawan_2006smqw.confE..52D}
{Dhawan}, V., {Mioduszewski}, A., and {Rupen}, M., Proceedings of
the VI Microquasar Workshop: Microquasars and
  Beyond. September 18-22, 2006, Como, Italy., p.52.1

\bibitem{Dubus:2006}
{Dubus}, G.,\aap, 2006, {\bf 456}, 801

\bibitem{Gupta_2006ApJ...650L.123G}
{Gupta}, S. and {B{\"o}ttcher}, M.,\apjl, 2006, {\bf 650}, L123

\bibitem{2010ATel_Veritas}
{Ong}, R.~A.: 2010, The Astronomer's Telegram {\bf 2948}, 1

\bibitem{EGRET_disc_1997ApJ...486..126K}
{Kniffen}, D.~A. et~al.,\apj, 1997, {\bf 486}, 126

\bibitem{AGILE_discovery_2009A&A...506.1563P}
{Pittori}, C. et~al.,\aap, 2009, {\bf 506}, 1563

\bibitem{Rolke:2004mj}
Rolke, W., Lopez, A., and Conrad, J., Nucl. Instrum. Meth., 2005,
{\bf A551}, 493

\bibitem{Romero_2005ApJ...632.1093R}
{Romero}, G.~E., {Christiansen}, H.~R., and {Orellana}, M., \apj,
2005, {\bf 632}, 1093

\bibitem{Tavani}
{Tavani}, M. et~al.,\apjl, 1998, {\bf 497}, L89

\bibitem{2011A&A...527A...9Z}
{Zabalza}, V., {Paredes}, J.~M., and {Bosch-Ramon}, V.,\aap, 2011,
{\bf 527}, A9+

\bibitem{Zdziarski_2010MNRAS.403.1873Z}
{Zdziarski}, A.~A., {Neronov}, A., and {Chernyakova}, M., \mnras,
2010, {\bf 403}, 1873

\end{thebibliography}
\end{document}